\RequirePackage{lineno}
\documentclass[final,5p,times,twocolumn]{elsarticle}

\usepackage{atlasphysics}
\usepackage{subfigure}
\usepackage{mathrsfs}
\usepackage{graphicx}\usepackage{hyperref}
\usepackage{multirow}
\usepackage{rotating}
\usepackage{url}
\usepackage{textpos}
\usepackage{color}
\usepackage{epstopdf}

\biboptions{sort,merge}

\newcommand{\meff}{\ensuremath{m_{\mathrm{eff}}}} 

\newcommand{\mttwo}{\ensuremath{m_\mathrm{T2}}}

\def\lsim{\mathrel{\rlap{\lower4pt\hbox{\hskip1pt$\sim$}}
    \raise1pt\hbox{$<$}}}                
\def\gsim{\mathrel{\rlap{\lower4pt\hbox{\hskip1pt$\sim$}}
    \raise1pt\hbox{$>$}}}                

\makeatletter
\renewcommand{\p@subfigure}{\arabic{figure}}
\renewcommand{\thesubfigure}{\alph{subfigure}}
\renewcommand{\@thesubfigure}{(\thesubfigure)}
\makeatother

\newcommand{\zeroleptitle}{Search for squarks and gluinos using final states with jets and missing transverse momentum with the ATLAS detector in $\sqrt{s}=7\TeV$ proton-proton collisions}

\newcommand{\handsomeAndDashingCorrespondingAuthors}[1]{}

\handsomeAndDashingCorrespondingAuthors{C.G. Lester, lester@hep.phy.cam.ac.uk}
\handsomeAndDashingCorrespondingAuthors{A.J. Barr,   a.barr@physics.ox.ac.uk}
\handsomeAndDashingCorrespondingAuthors{S. Caron,    scaron@physik.uni-freiburg.de}

\newcommand{\maybesection}[1]{\par  \vspace{1.6mm} {{\bf #1}} \vspace{0.2mm}\par}
\newcommand{\mytabref}[1]{Table~\ref{#1}}
\newcommand{\myfigref}[1]{Figure~\ref{#1}}
\newcommand{\regionmttwo}{\mathrm{B}}
\newcommand{\regionmeffjj}{\mathrm{A}}
\newcommand{\regionmeffjjjbig}{\mathrm{D}}
\newcommand{\regionmeffjjjsmall}{\mathrm{C}}

\newcommand{\anactualptmissvecwithapinit}{{\vec{P}}_\mathrm{T}^\mathrm{miss}}

\newcommand{\ourvecptmiss}{\ensuremath{\anactualptmissvecwithapinit}}
\newcommand{\ourmagptmiss}{\MET}

\newcommand{\ourdeltaphishort}{\Delta\phi}
\newcommand{\ourdeltaphifull}{\ourdeltaphishort(\textrm{jet},\ourvecptmiss)_\mathrm{min}}

\newcommand{\ttv}[2]{{\tt {#1}{#2}}}
\newcommand{\alpgen}{{\tt ALPGEN}}

\newcommand{\prospino}{{\tt PROSPINO}}
\newcommand{\herwig}{{\tt HERWIG}}
\newcommand{\pythia}{{\tt PYTHIA}}
\newcommand{\mcatnlo}{{\tt MC@NLO}}
\newcommand{\fewz}{{\tt FEWZ}}

\newcommand{\isajet}{{\tt ISAJET}}

\newcommand{\isasusy}{{\tt ISASUSY}}
\newcommand{\jimmy}{{\tt JIMMY}}
\newcommand{\cteqVER}[1]{{\tt CTEQ#1}}
\newcommand{\mrstVER}[1]{{\tt MRST#1}}
\newcommand{\ourintlumi}{{$35~\ipb$}}
\newcommand{\ourpt}{p_\mathrm{T}}
\newcommand{\note}{letter}
\newcommand{\mSUGRA}{MSUGRA/CMSSM}
\newcommand{\mSUGRAShortened}{CMSSM}
\newcommand{\citemsugraandcmssm}{\cite{Chamseddine:1982jx,*Barbieri:1982eh,*Ibanez:1982ee,*Hall:1983iz,*Ohta:1982wn,*Kane:1993td}}

\newcommand{\degenerateMassLimitSqGlPlane}{870}
\newcommand{\degenerateMassLimitSugraPlane}{775}

\newcommand{\absolteGluinoMassLimitSqGlPlane}{500}

\newcommand{\abstractLimitsNugget}{Gluino masses below \absolteGluinoMassLimitSqGlPlane~\GeV{} are excluded at the 95\% confidence level in simplified models containing only squarks of the first two generations, a gluino octet and a massless neutralino. The exclusion increases to \degenerateMassLimitSqGlPlane~\GeV{} for equal mass squarks and gluinos. In \mSUGRA{} models with $\tan\beta=3$, $A_0=0$ and $\mu> 0$, squarks and gluinos of equal mass are excluded below \degenerateMassLimitSugraPlane~\GeV. These are the most stringent limits to date.}

\begin{document}
\begin{frontmatter}
\title{\zeroleptitle}
\author{The ATLAS Collaboration}

\begin{abstract}
\begin{textblock}{6}[0,1](0,-3){\small CERN-PH-EP-2011-022, Submitted to Phys. Lett. B}
\end{textblock}
A search for squarks and gluinos in final states containing jets, missing
transverse momentum and no electrons or muons is presented. 
The data were recorded by the ATLAS experiment in $\sqrt{s}=7\TeV$ proton-proton collisions at the Large Hadron Collider.
No excess above the Standard Model background expectation was observed in \ourintlumi{} of analysed data.  \abstractLimitsNugget  
\end{abstract}
\end{frontmatter}


\maybesection{1. Introduction}

Many extensions of the Standard Model (SM) include heavy coloured
particles, some of which could be accessible at the LHC.  The squarks and
gluinos of supersymmetric theories \cite{susy} are one example of such
particles.  This \note{} presents the first ATLAS search for squarks
and gluinos in final states containing only jets and large missing transverse momentum.
Interest in this final state is motivated by the large number of $R$-parity conserving
models in which squarks and gluinos can be produced in pairs \{$\gluino\gluino$, 
$\squark \squark$, $\squark
\gluino$\} and can generate that final state in their decays
$\squark \rightarrow q
\ninoone$
and $\gluino
\rightarrow q q \ninoone$ to weakly interacting neutralinos, $\ninoone$, which escape the detector
unseen.
The analysis presented here is based on a study of purely hadronic
final states; events with reconstructed electrons and muons are
vetoed to avoid overlap with a related ATLAS search \cite{Collaboration:2011hh} which requires a lepton.
The search strategy was optimised for
maximum exclusion in the $(m_{\gluino},m_{\squark})$-plane for a set
of simplified models in which all other supersymmetric particles (except for
the lightest neutralino) were given masses beyond the reach of the LHC.
Though interpreted in terms of supersymmetric models, the main results of this analysis (the data and expected background event counts in the signal regions) are relevant for excluding any model of new physics that predicts jets in association with missing transverse momentum.  Currently, the most stringent limits on squark and gluino masses are obtained 
at the LHC~\cite{Khachatryan:2011tk} and at the
Tevatron~\cite{Abachi:1995ng,*Affolder:2001tc,*Aaltonen:2008rv,*:2007ww,*Abazov:2009zi}
.

\maybesection{2. The ATLAS Detector and Data Samples}

The ATLAS detector~\cite{Aad:2008zzm} is a multipurpose particle
physics apparatus with a forward-backward symmetric cylindrical
geometry and nearly 4$\pi$ coverage in solid angle.\footnote{
ATLAS uses a right-handed coordinate system with its origin at the nominal
interaction point in the centre of the detector and the $z$-axis along the
beam pipe. Cylindrical coordinates $(r,\phi)$ are used in the transverse
plane, $\phi$ being the azimuthal angle around the beam pipe. The pseudorapidity $\eta$ is
defined in terms of the polar angle $\theta$ by $\eta=-\ln\tan(\theta/2)$.
}  The layout of the detector is
dominated by four superconducting magnet systems, which comprise a
thin solenoid surrounding inner tracking detectors and three large
toroids supporting a large muon tracker.
The calorimeters are of particular importance to this analysis.  In the pseudorapidity
region $\left|\eta\right| < 3.2$, high-granularity liquid-argon (LAr)
electromagnetic (EM) sampling calorimeters are used.  An iron-scintillator tile
calorimeter provides hadronic coverage over
$\left|\eta\right| < 1.7$.  The end-cap and forward regions, spanning
$1.5 < \left|\eta\right| < 4.9$, are instrumented with LAr calorimetry
for both EM and hadronic measurements.

The data sample used in this analysis was taken in 2010 with the LHC operating at a centre-of-mass energy of 7~\TeV.
Application of beam, detector and data-quality requirements resulted in
a total integrated luminosity of \ourintlumi.
The detailed
trigger specification varied throughout the data-taking period, partly as a consequence of the rapidly
increasing LHC luminosity, but always
guarantees a trigger efficiency above 97\% for events with  a reconstructed jet with transverse momentum ($\ourpt$) exceeding $120 \GeV$ and more than 100~\GeV{} of missing $\ourpt$.

\maybesection{3. Object Reconstruction}

Jet candidates are reconstructed by using the
anti-$k_t$ jet clustering algorithm~\cite{Cacciari:2008gp,*Cacciari:2005hq} with a
distance parameter of $0.4$. The inputs to this algorithm are clusters
of calorimeter cells seeded by those with energy significantly above
the measured noise. Jet momenta are constructed by performing a
four-vector sum over these cell clusters, treating each as an
$(E,\vec{p})$ four-vector with zero mass.  These jets are corrected for
the effects of calorimeter non-compensation and inhomogeneities by
using $\ourpt$- and $\eta$-dependent calibration factors
based on Monte Carlo (MC) corrections validated with extensive
test-beam and collision-data studies \cite{atlas-jetcal}. Only jet candidates with $\ourpt > 20$
\GeV{} and $|\eta| < 4.9$ are retained hereafter.

Electron candidates are required to have $\ourpt > 10~\GeV$, to have $|\eta| <
2.47$, to pass the `medium' electron shower shape and track selection criteria of
Ref.~\cite{Aad:2010yt}, and to be outside problematic regions of the calorimeter.
Muon candidates are required to have $\ourpt > 10$ \GeV{} and $|\eta| <
2.4$.  The sum of the transverse momenta of charged particle tracks within a cone of radius
$\Delta R = \sqrt{(\Delta \eta)^2 +
(\Delta \phi)^2} = 0.2$ around the muon trajectory is required to be less
than $1.8$ \GeV.

Following the steps above, overlaps between candidate jets with $|\eta|<2.5$ and leptons are resolved using the method of Ref.~\cite{atlas-csc} as follows.
First, any such jet candidate lying within a distance $\Delta
R<0.2$ of an electron is discarded. Then the whole event is rejected
if any  electron candidate remains in the calorimeter
transition region $1.37 < |\eta| < 1.52$ between barrel and
end-cap. Finally, any lepton candidate remaining within a distance
$\Delta R =0.4$ of such a jet candidate is discarded. 

The measurement of the missing transverse momentum two-vector
$\ourvecptmiss$ (and its magnitude $\ourmagptmiss$) is then based on
the transverse momenta of all remaining jet and lepton candidates and all
calorimeter clusters not associated to such objects.  Following this, all jet candidates with $|\eta|>2.5$ are discarded. 

Thereafter, the remaining lepton and jet candidates are considered
``reconstructed'', and the term ``candidate'' is dropped.

\maybesection{4. Event Selection}\label{sec:eventselection}

Following the object reconstruction described above, events are
discarded if any electrons or muons
remain, or if they have any jets
failing quality selection criteria against detector noise and
against non-collision backgrounds
\cite{atlas-jet-clean}, or if they lack a reconstructed primary
vertex associated with five or more tracks.

In order to achieve maximal reach over the
$(m_{\gluino},m_{\squark})$-plane, several signal regions are defined.
When production of squark pairs $\squark \squark$ is dominant, only
a small number of jets (one per squark from $\squark \rightarrow q
\ninoone$) is expected. The optimal strategy for the $\squark\squark$ region
therefore makes requirements on two jets only.  When production
involves gluinos ($\gluino\gluino$ and $\squark\gluino$), extra jets are expected from $\gluino\rightarrow q q \ninoone$.  In these regions,
requiring at least three jets yields better sensitivity.
The higher total cross section in the associated $\squark\gluino$
region where both species are accessible permits the use of tighter
criteria than in the $\gluino\gluino$ region. Four signal regions 
$\regionmeffjj$,
$\regionmttwo$,
$\regionmeffjjjsmall$ and
$\regionmeffjjjbig$
are thus defined (targeting
light-$\squark\squark$,
heavy-$\squark\squark$,
$\gluino\gluino$ and
$\gluino\squark$
production, respectively) as shown in \mytabref{tab:signal_regions}.  In this table, $\ourdeltaphifull$ is the smallest of the
azimuthal separations between $\ourvecptmiss$ and jets with $\ourpt>40~\GeV$ (up to a maximum of three, in descending order of $\ourpt$,
whether pre-selected or not).  The variable \mttwo{} \cite{Lester:1999tx,*Barr:2003rg,*Cheng:2008hk} is defined 
to be the maximal lower bound on
the mass of a pair produced particle which decays into one
of the pre-selected jets and a massless undetected particle, assuming
the two undetected particles are the only source of the event $\ourvecptmiss$.
The effective mass, \meff, is defined as the sum of $\ourmagptmiss$ and the magnitudes of the transverse momenta of the two highest \pT{} jets (in signal region $\regionmeffjj$) or three highest \pT{} jets (in signal regions $\regionmeffjjjsmall$ and $\regionmeffjjjbig$).
The $\squark\squark$ channel has two signal regions, $\regionmeffjj$ and $\regionmttwo$, because the \mttwo{} distribution has the best expected reach in $m_{\squark}$, but \meff{} offers better coverage for lighter squarks. 

\newcommand\ourtwidth{2.0cm}
\newcommand\yourtwidth{1.5cm}
\newcommand\errorSTAT{(m)}
\newcommand\errorUNCORR{[u]}
\newcommand\errorUNCORRANDJES{[u+j]}
\newcommand\errorJES{[j]}
\newcommand\errorJER{(r)}
\newcommand\errorLUMI{[{$\mathcal L$}]}
\newcommand\errorACC{(a)}
\newcommand\errorLE{($\epsilon$)}
\newcommand\errorCR{(c)}
\newcommand\errorEXT{(x)}
\newcommand\errorCREXT{(c+x)}

\begin{table}
  \begin{center}\renewcommand\arraystretch{1.2}
\renewcommand\tabcolsep{1.2mm}
    \begin{tabular}{llcccc}
\hline
      \hline
 & 
& $\regionmeffjj$ & $\regionmttwo$
& $\regionmeffjjjsmall$ & $\regionmeffjjjbig$ 
\\ 
\hline
\multirow{4}{3mm}{\rotatebox{90}{{\small \mbox{Pre-selection}}}}
& Number of required jets   & $\ge 2$  & $\ge 2$ & $\ge 3$ & $\ge 3$ \\
& Leading jet $\ourpt$ [\GeV]   & $>120$  & $>120$ & $>120$ & $>120$ \\ 
& Other jet(s) $\ourpt$ [\GeV]   & $>40$ &  $>40$ &  $>40$ &  $>40$ \\ 
& $\ourmagptmiss$ [\GeV]  & $>100$  & $>100$  & $>100$  & $>100$ \\ 
\hline
\multirow{4}{3mm}{\rotatebox{90}{{\small \mbox{Final selection}}}}
& $\ourdeltaphifull$ & $>0.4$  & $>0.4$  & $>0.4$  & $>0.4$ \\ 
& $\ourmagptmiss/\meff$  & $>0.3$ & -- & $> 0.25$ & $> 0.25$ \\ 
& $\meff$ [\GeV] & $>500$ &  -- & $>500$ & $>1000$ \\ 
& $\mttwo$ [\GeV] & -- & $>300$ & -- & --  \\ 
\hline
\hline
\end{tabular}
\caption{Criteria for admission to each of the four overlapping signal regions A to D.
 All variables are defined in \S4
.}
    \label{tab:signal_regions}
  \end{center}
\end{table}

\begin{table*}\begin{center}\renewcommand\arraystretch{1.1}

\begin{tabular*}{\textwidth}{l@{\ \ \ }l@{\ \ \ \ }l@{\ \ \ \ }l@{\ \ \ \ }l}
\hline\hline
               & $\phantom{8}$Signal region $\regionmeffjj$
               & Signal region $\regionmttwo$
               & Signal region $\regionmeffjjjsmall$ 
               & Signal region $\regionmeffjjjbig$
               \\
\hline
QCD
			& $\phantom{88}7 \;{}^{+8}_{-7} \textrm{\errorUNCORRANDJES}$
			& $\phantom{8}0.6 \;{}^{+0.7}_{-0.6} \textrm{\errorUNCORRANDJES}$
			& $\phantom{8}9 \;{}^{+10}_{-9} \textrm{\errorUNCORRANDJES}$
			& $0.2 \;{}^{+0.4}_{-0.2} \textrm{\errorUNCORRANDJES}$
				\\
$W$+jets
 		& $\phantom{8}50 \pm 11  \textrm{\errorUNCORR}
					\;{}^{+14}_{-10} \textrm{\errorJES} \pm \phantom{8}5\textrm{\errorLUMI}
				$ 
		& $\phantom{8}4.4 \pm 3.2  \textrm{\errorUNCORR}
					\;{}^{+1.5}_{-0.8} \textrm{\errorJES} \pm 0.5\textrm{\errorLUMI}
				$
 		& $35 \pm \phantom{8}9 \textrm{\errorUNCORR}
					\;{}^{+10}_{-\phantom{8}8} \textrm{\errorJES}\pm 4\textrm{\errorLUMI}
				$ 
		& $1.1 \pm 0.7 \textrm{\errorUNCORR}
					\;{}^{+0.2}_{-0.3} \textrm{\errorJES} \pm 0.1\textrm{\errorLUMI}
				$
				\\
$Z$+jets
		& $\phantom{8}52 \pm 21  \textrm{\errorUNCORR} 
					\;{}^{+15}_{-11} \textrm{\errorJES} \pm \phantom{8}6\textrm{\errorLUMI}
				$ 
		& $\phantom{8}4.1 \pm 2.9  \textrm{\errorUNCORR} 
					 \;{}^{+2.1}_{-0.8} \textrm{\errorJES} \pm 0.5\textrm{\errorLUMI}
				$
	 	& $27 \pm 12  \textrm{\errorUNCORR}
					\;{}^{+10}_{-\phantom{8}6} \textrm{\errorJES}\pm 3\textrm{\errorLUMI}
				$
		& $0.8 \pm 0.7  \textrm{\errorUNCORR} 
					\;{}^{+0.6}_{-0.0} \textrm{\errorJES}\pm 0.1\textrm{\errorLUMI}
				$
				\\
\ttbar{} and $t$
 	& $\phantom{8}10 \pm \phantom{8}0  \textrm{\errorUNCORR}
			\;{}^{+\phantom{8}3}_{-\phantom{8}2} \textrm{\errorJES}\pm \phantom{8}1\textrm{\errorLUMI}$
	& $\phantom{8}0.9 \pm 0.1  \textrm{\errorUNCORR}
			\;{}^{+0.4}_{-0.3} \textrm{\errorJES} \pm 0.1\textrm{\errorLUMI}$
 	& $17 \pm \phantom{8}1  \textrm{\errorUNCORR}
			\;{}^{+\phantom{8}6}_{-\phantom{8}4} \textrm{\errorJES}\pm 2\textrm{\errorLUMI}$
	& $0.3 \pm 0.1  \textrm{\errorUNCORR}
			\;{}^{+0.2}_{-0.1} \textrm{\errorJES} \pm 0.0\textrm{\errorLUMI}$ 
			       \\
\hline
Total SM
                & $118 \pm 25\textrm{\errorUNCORR}
					\;{}^{+32}_{-23} \textrm{\errorJES} \pm 12\textrm{\errorLUMI}
					          $
			& $10.0 \pm 4.3 \textrm{\errorUNCORR}
					 \;{}^{+4.0}_{-1.9} \textrm{\errorJES} \pm 1.0\textrm{\errorLUMI}
					          $
                & $88 \pm 18 \textrm{\errorUNCORR} 
					\;{}^{+26}_{-18} \textrm{\errorJES}\pm 9\textrm{\errorLUMI}
					           $
		& $2.5 \pm 1.0 \textrm{\errorUNCORR} 
					 \;{}^{+1.0}_{-0.4} \textrm{\errorJES} \pm 0.2\textrm{\errorLUMI}
					         $
				\\ 
Data  & $\phantom{8}87$ & $11$ & $66$ & $2$ \\
\hline
\hline
\end{tabular*}

\caption{\label{tab:Background_allDD_35pb} 
Expected and observed numbers of events in the four signal regions.
Uncertainties shown are due to ``MC statistics, statistics in control regions, other sources of uncorrelated systematic uncertainty, and also the jet energy resolution and lepton efficiencies'' \errorUNCORR,  the jet energy scale \errorJES, and the luminosity \errorLUMI. 
\label{tab:expectedEvents} }
\end{center}\end{table*}

\begin{figure*}
  \centering
    \includegraphics[width= 0.485\textwidth]{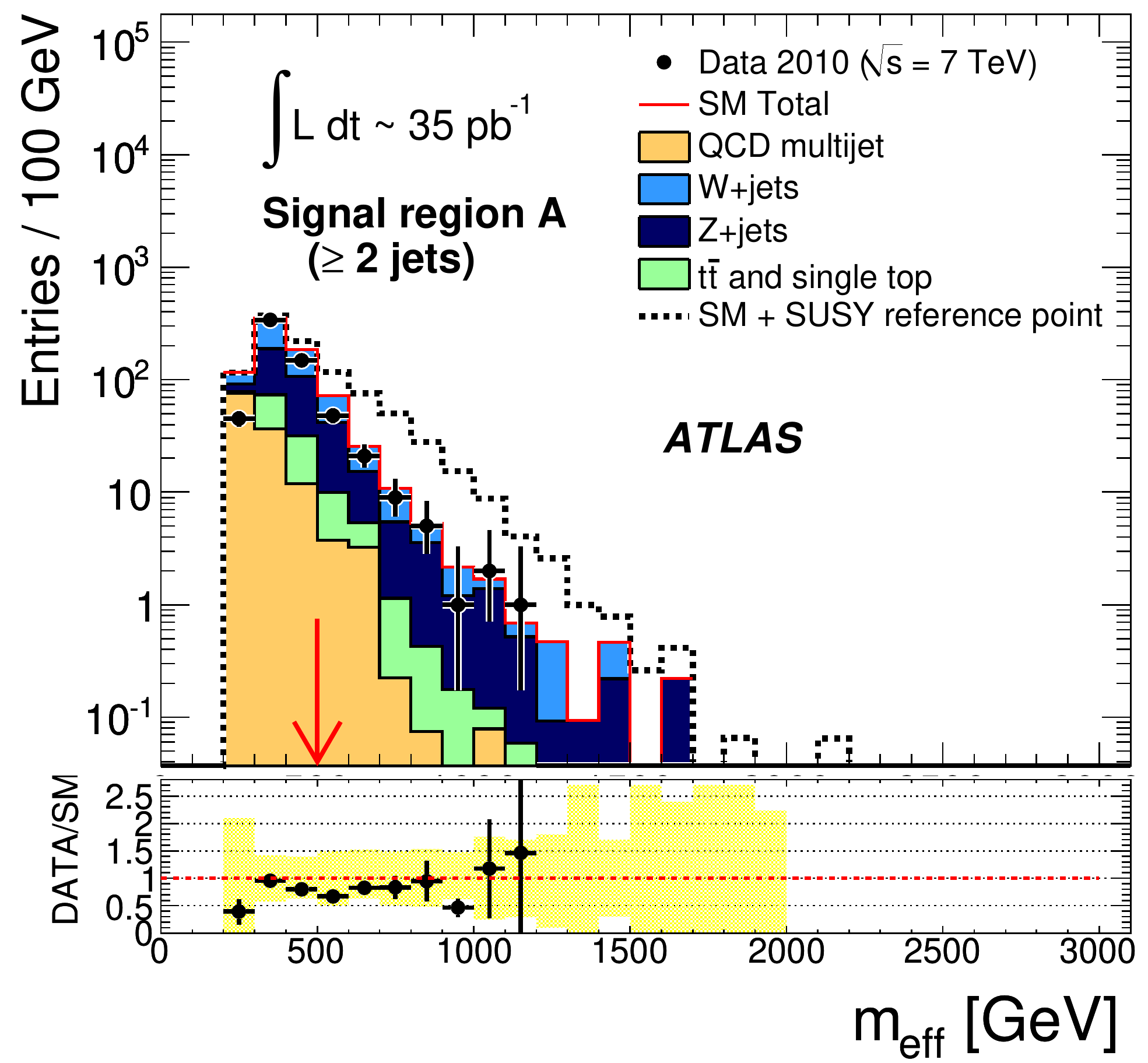}\includegraphics[width= 0.485\textwidth]{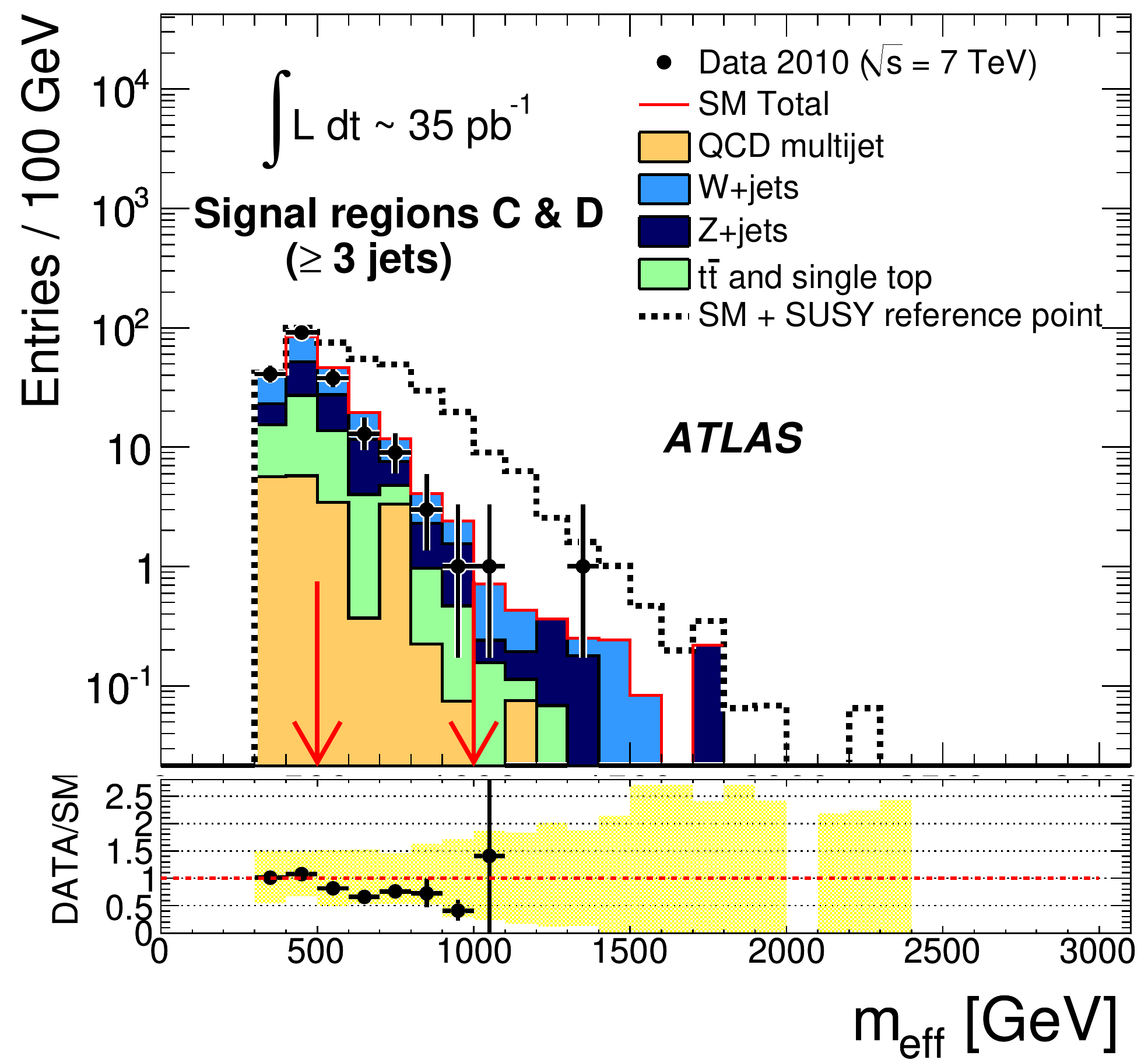}
    \includegraphics[width=0.485\textwidth]{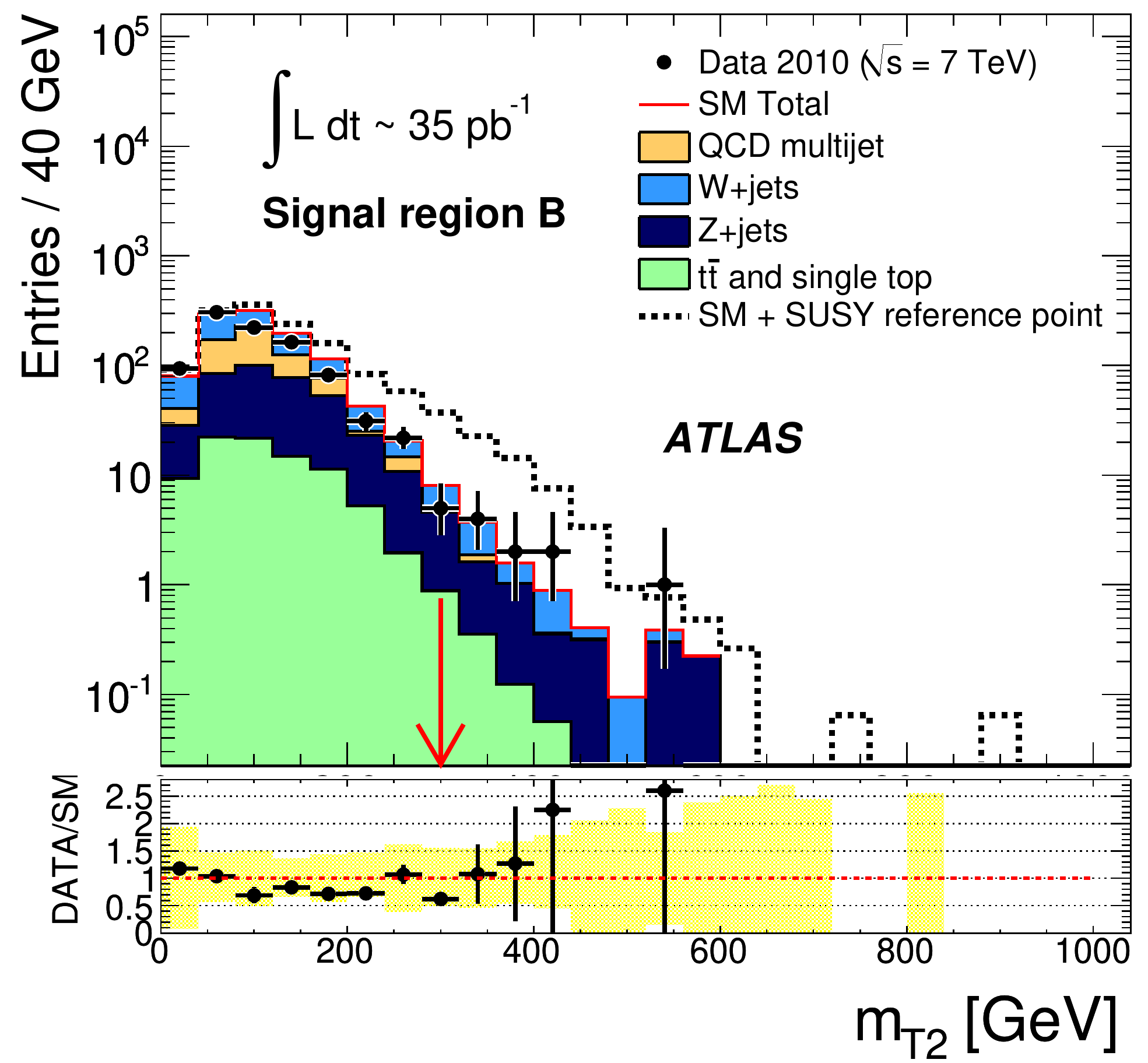}\includegraphics[width=0.485\textwidth]{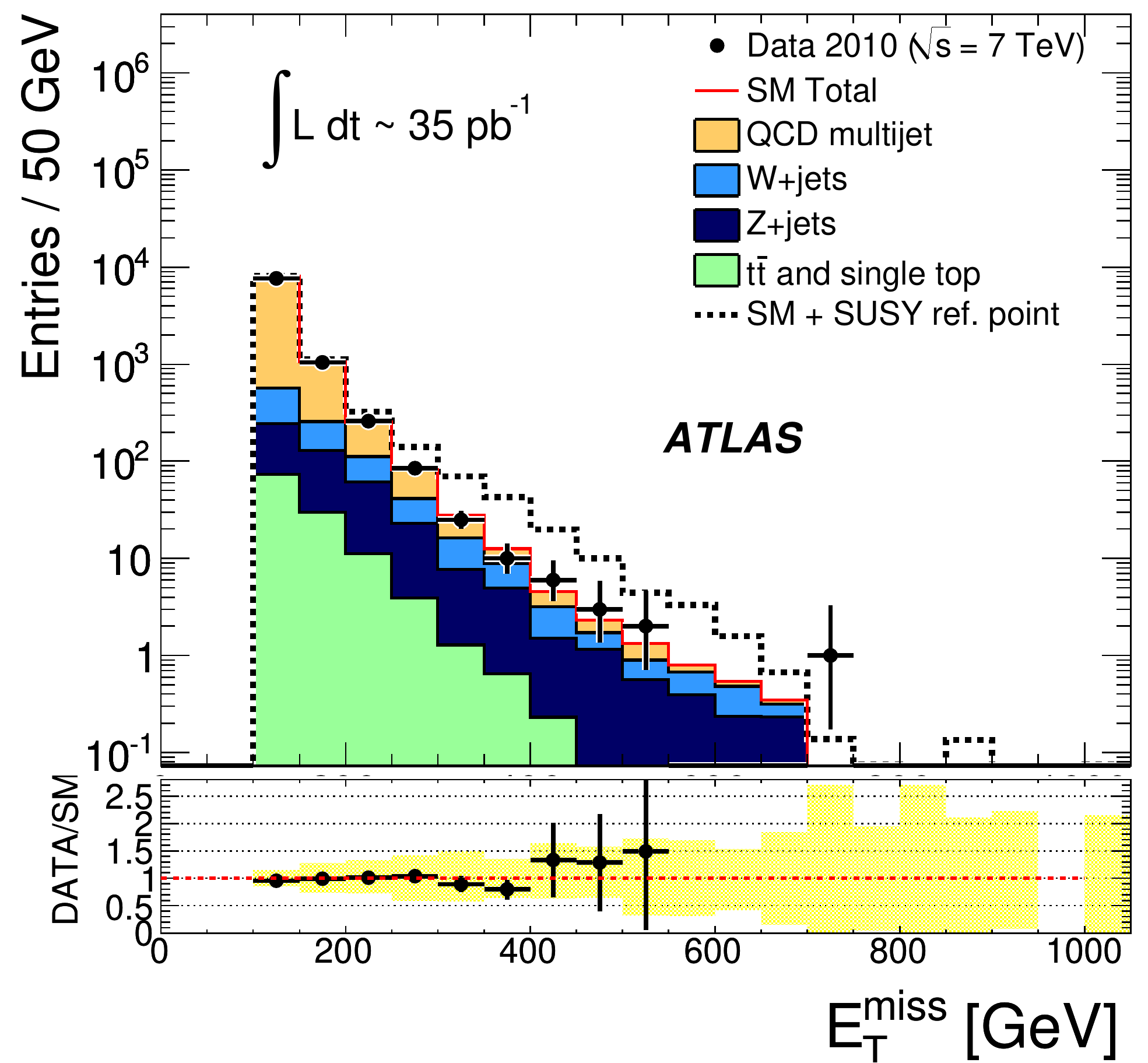}
  \caption{The distributions of \meff{} (separately for the $\ge 2$ and $\ge 3$ jet regions) and \mttwo{} are shown for data and for the expected SM contributions after application of all selection criteria -- cuts on the variables themselves are indicated by the red arrows.  Also shown is the $\ourmagptmiss$ distribution after the $\ge$2~jet preselection cuts only.  For comparison, each plot includes a curve showing the expectation for an \mSUGRA{} reference point with  $m_0=200$~\GeV, $m_{1/2}=190$~\GeV, $A_0=0$, $\tan\beta=3$ and $\mu> 0$.  This reference point is also indicated by the star on \myfigref{fig:sugraExclusion}.  Below each plot may be seen the ratio of the data to the SM expectation.  Black vertical bars show the statistical uncertainty from the data, while the yellow band shows the size of the Standard Model MC uncertainty.
  \label{fig:some_actual_plots}}
\end{figure*}


\maybesection{5. Backgrounds, Simulation and Normalisation}

Standard Model background processes contribute to the event counts in
the signal regions. The dominant sources are $W+$jets, $Z+$jets, top
pair, QCD multi-jet (hereafter, the expression ``multi-jet'' is dropped) and single top production.
Non-collision backgrounds are negligible.
The majority of the $W$+jets background is composed 
of $W\rightarrow\tau\nu$ events, or $W\rightarrow l\nu$ events in which  
no electron or muon candidate is reconstructed.
The largest part of the $Z$+jets background comes from the irreducible
component in which $Z\rightarrow\nu\bar\nu$ generates large $\ourmagptmiss$. 
Hadronic $\tau$ decays in $t \bar t \rightarrow b \bar b \tau \nu q q $
can generate large $\ourmagptmiss$ and pass the jet
and lepton requirements at a non-negligible rate.
The QCD background in the signal regions is predominantly caused by
poor reconstruction of jet energies in calorimeters leading to `fake'
missing transverse momentum.  There is also a
contribution from neutrinos when events
contain leptonic decays of heavy quarks.
Extensive validation of MC against data has been performed for each of
these background sources and for a wide variety of control
regions. The excellent agreement found motivates an approach in which
both the shape and the normalisation of the $W$+jets, $Z$+jets and top
backgrounds are taken from MC simulation. In contrast, the
QCD background is normalised to data in
control regions as described below.

\par

Production of $W$ and $Z$ bosons, in association with jets, was
simulated with \alpgen{} \cite{Mangano:2002ea} v2.13 at leading
order (LO) and up to $2\rightarrow 5$ partons using \cteqVER{6L1}{}
PDFs~\cite{Pumplin:2002vw}.  Both were separately normalised to the
next-to-next-to-leading-order inclusive $W$ and $Z$ cross sections
from \fewz{} \cite{Melnikov:2006kv,*Gavin:2010az} v2.0.  Both
resulting samples were found to be consistent with a variety of
data-derived estimates, including methods based on: re-simulation of reconstructed leptons as hadronically decaying
taus; removal of leptons from $W(l\nu)$+jet and $Z(ll)$+jet events;
and by comparing MC predictions to data in control regions enriched
with background events.

Production of top quarks (both singly and in pairs, assuming
$m_\mathrm{top}=172.5$ \GeV) was simulated with \mcatnlo{}
\cite{Frixione:2002ik,*Frixione:2003ei} {v3.41} using \cteqVER{6.6}{} next-to-leading-order
(NLO) PDFs \cite{Nadolsky:2008zw}.  This estimate was found to be
consistent with a data-driven cross-check based on
replacement of reconstructed muons in the corresponding
single lepton channels with simulated hadronic $\tau$ decays.  Agreement was also found after reweighting
the $t\bar t$ MC according to experimentally measured $b$-tag weights.

Simulated QCD events were generated both with
\pythia{} \cite{Sjostrand:2006za} {v6.4.21}, which uses
$2\rightarrow 2$ LO matrix elements (ME) with the
\mrstVER{2007 LO*}{} PDF set \cite{Sherstnev:2007nd}, and with \alpgen{}
implementing the exact LO ME for up to
$2\rightarrow 5$ partons.  The normalisation of these samples was
fixed by a scaling designed to achieve a match to data in control
regions obtained by reversing the $\ourdeltaphishort$ requirements.
After this scaling, both sets of simulations were in
agreement within the experimental uncertainties, and therefore only \pythia{}
QCD simulations are used further in this analysis.  
The resulting QCD simulation was found to be consistent with a data-driven QCD estimate in which high $\ourmagptmiss$ events
were generated from data by smearing low $\ourmagptmiss$ events on a
jet-by-jet basis with measured jet energy resolution functions. This latter technique has no MC dependencies; it provides a completely
independent determination of the QCD background using only quantities measured
from the data. Additional control regions having reversed $\ourmagptmiss/\meff$ requirements were used as further checks on the normalisation.

Supersymmetric events were generated with \ttv{\herwig}{++}{}
\cite{Bahr:2008pv} {v2.4.2}.
These samples were normalised
using NLO cross sections determined by \prospino{}
\cite{Beenakker:1996ch} {v2.1}.

All non-\pythia{} samples used \ttv{\herwig}{++}{} or \ttv{\herwig}{-6.510} \cite{Corcella:2000bw} to simulate parton showering and fragmentation, while \jimmy{}{} \cite{Butterworth:1996zw} {v4.31} was used to generate the underlying event.  All samples were produced using an ATLAS `tune' \cite{mc09tunes} and a full detector simulation \cite{:2010wqa}.

\maybesection{6. Systematic Uncertainties}

\begin{figure}
  \centering
  \includegraphics[width=0.97\columnwidth]{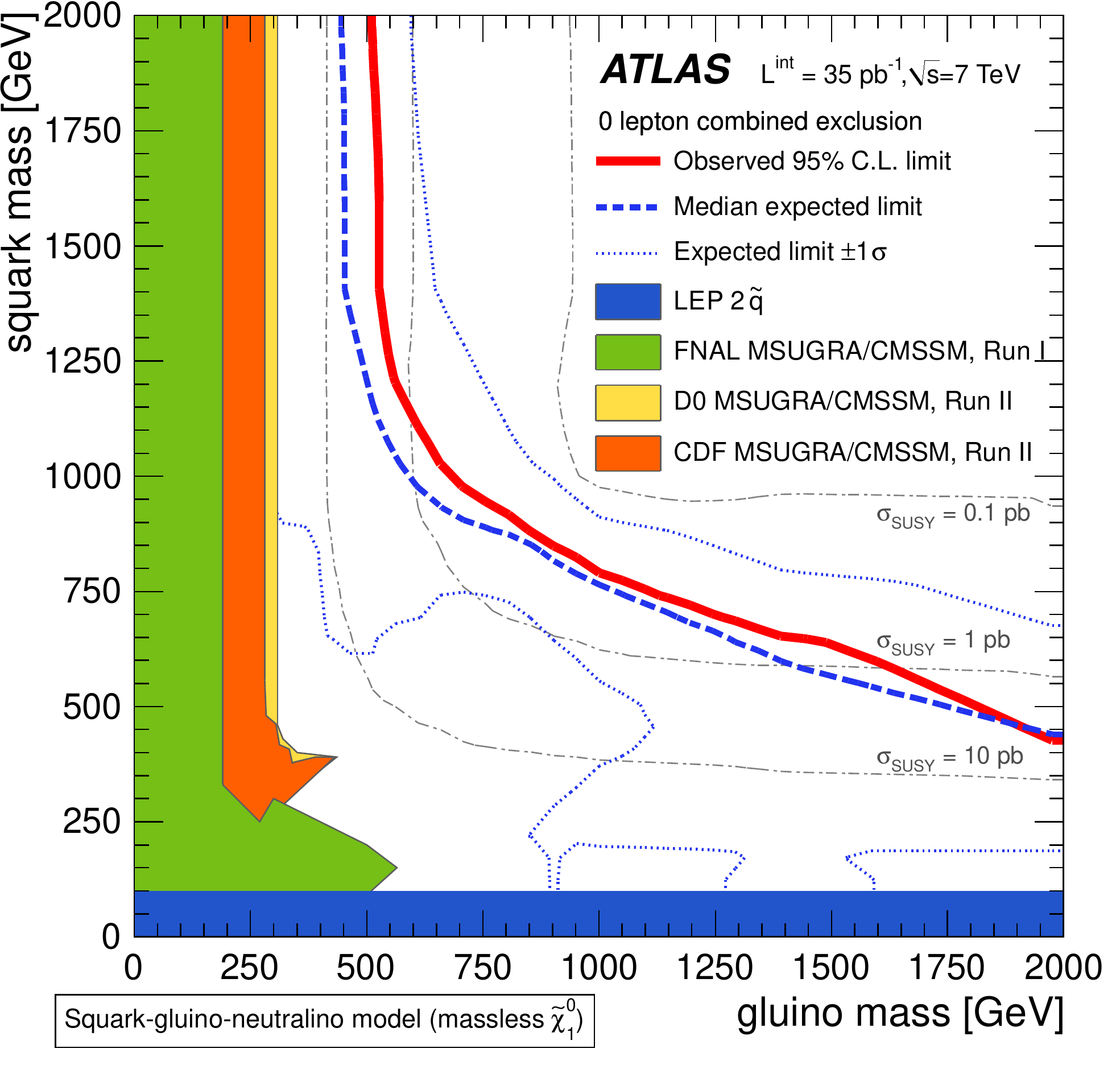}
\caption{95\% C.L. exclusion limits in the ($m_{\gluino}$, $m_{\squark}$) plane together with existing limits \cite{Abachi:1995ng,Affolder:2001tc,Aaltonen:2008rv,:2007ww,Abazov:2009zi}.   Comparison with existing limits is illustrative only as some are derived in the context of \mSUGRA{} or may not assume $m_{\ninoone}=0$.
  \label{fig:squarkGluinoExclusion}}
\end{figure}

\begin{figure}
  \centering
  \includegraphics[width=0.97\columnwidth]{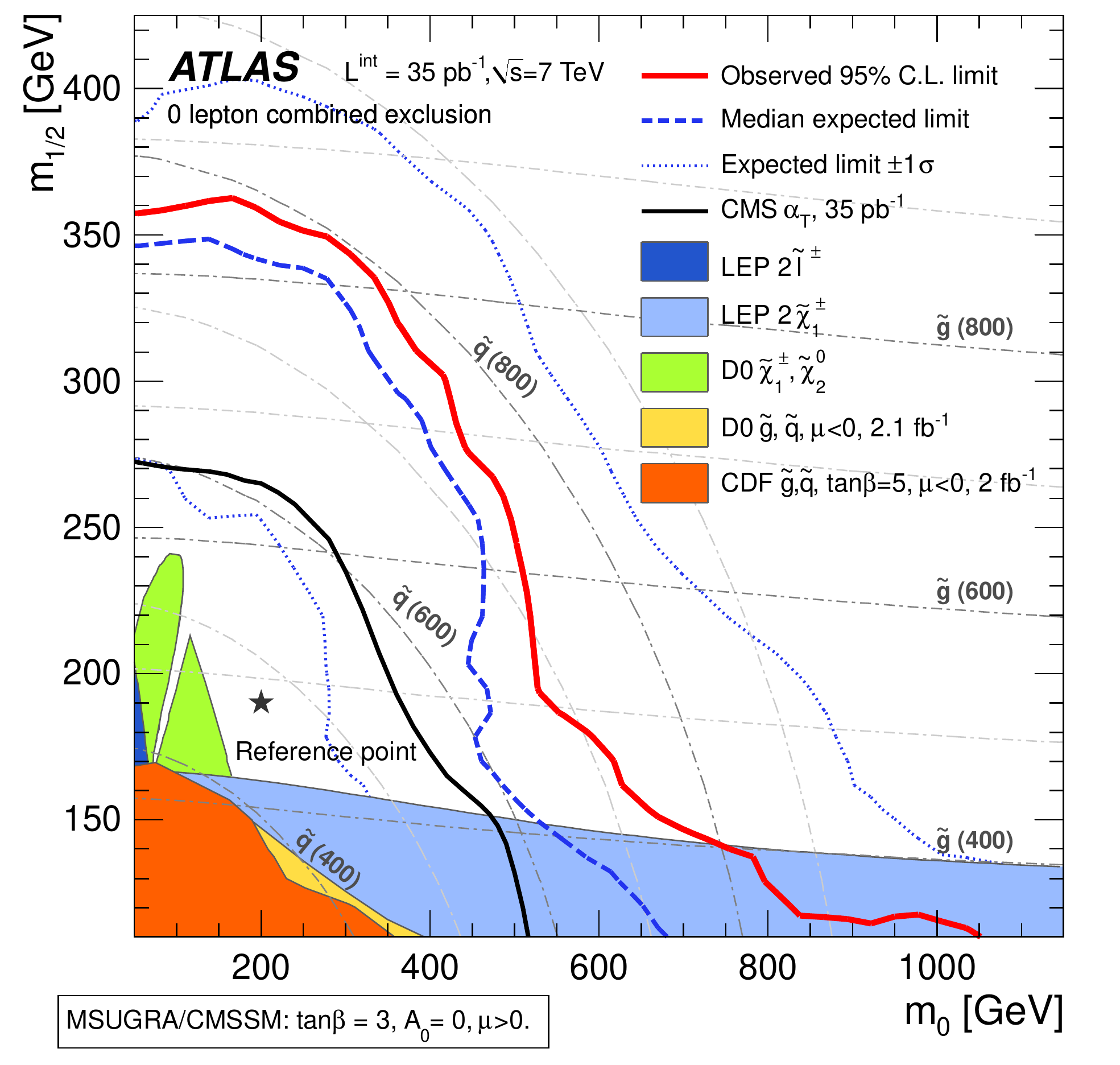}
\caption{95\% C.L. exclusion limits in the $\tan\beta=3$, $A_0=0$ and $\mu>0$ slice of \mSUGRA, together with existing limits \cite{Aaltonen:2008rv,:2007ww,Abazov:2009zi,Khachatryan:2011tk} with the different model assumptions given in the legend.
  \label{fig:sugraExclusion} }
\end{figure}

The primary sources of systematic uncertainties in the background
estimates are: the jet energy scale (JES),
the jet energy resolution
(JER), the luminosity determination, the MC modelling, the lepton efficiencies, the
extrapolation from control regions into signal
regions, and the finite statistics of the MC samples and control regions.
The uncertainty on the luminosity determination is estimated to be
11\%~\cite{atlas-lumi}.
The JES uncertainty has been
measured from the complete 2010 data set using the techniques
described in Ref.~\cite{atlas-jetcal} 
and, though $\ourpt$ and $\eta$ dependent, is around 7\%.
The JER measured in data \cite{atlas-jer} was applied to all MC
simulated jets and was propagated to $\ourvecptmiss$.  The difference
between the re-calibrated and nominal MC is taken as the systematic uncertainty due to
this effect.
The uncertainty on the estimated top background is dominated by the JES uncertainty.
Systematic uncertainties associated with mis-identification of leptons, jet
energy scale inter-calibration, the rate of leptonic $b$-decays and the
non-Gaussian tail of the jet response function have also been incorporated 
where appropriate.

Systematic uncertainties on the SUSY signal were estimated by variation
of the factorisation and renormalisation scales in \prospino{} between half
and twice their default values and by considering the PDF uncertainties provided by \cteqVER{6}.
Uncertainties were calculated for individual
production processes (e.g.~$\squark\squark$, $\gluino\gluino$, etc.).

%

\maybesection{7. Results, Interpretation and Limits}

The number of observed data events and the number of SM events expected to enter each of the signal regions are shown in \mytabref{tab:expectedEvents}. The background model is found to be in good agreement with the data, and the distributions of \meff{}, \mttwo{} and $\ourmagptmiss$ are shown in \myfigref{fig:some_actual_plots}.
%
%
%
%
\par
An interpretation of the results is presented in
\myfigref{fig:squarkGluinoExclusion}{} as a 95\% confidence exclusion
region in the $(m_{\gluino},m_{\squark})$-plane for the simplified set of
models with $m_{\ninoone}=0$ for which the analysis was optimised.  In these models the gluino mass and the masses of the
squarks of the first two generations are set to
the values shown in the figure. All other supersymmetric particles,
including the squarks of the third generation, are decoupled by being given masses of 5~\TeV. 
{
\isasusy{} from
\isajet{} \cite{Paige:2003mg} {v7.80} was used to calculate the
decay tables, and to guarantee consistent electroweak symmetry
breaking.}
The SUSY Les Houches
Accord files for the models used may be found online \cite{auxmaterialslhafiles}. 
The results are also interpreted in the
$\tan\beta=3$, $A_0=0$, $\mu>0$ slice of \mSUGRA~\citemsugraandcmssm{}
in \myfigref{fig:sugraExclusion}.  
\par
These figures also show the variation of the expected limit in response to $\pm 1\sigma$
fluctuations of the SM expectation including the stated systematic uncertainties.
The exclusion regions are constructed using a profile likelihood ratio method.  
Pseudo-experiments are used to compute one-sided upper limits on the signal contribution, which
is assumed to be non-negative. 
At each point in the $(m_{\gluino},m_{\squark})$ and \mSUGRAShortened{} planes, the chosen test statistic is the likelihood ratio corresponding to the number of observed events in the signal region whose expected sensitivity was largest.  Plots showing where each signal region is dominant may be found in \cite{auxmaterialplots}. All signal regions contribute to the exclusion and to its boundary in the $(m_{\gluino},m_{\squark})$-plane.   Region $\regionmeffjjjbig$ is dominant near the \mSUGRAShortened{} boundary.  
In the simplified model, changing the $\ninoone$ mass from 0 to 100~\GeV{} 
reduces the number of selected events by only $\lsim20\% $
near the exclusion curve so only slightly modifies the
excluded region in the $(m_{\gluino},m_{\squark})$-plane.
In the \mSUGRAShortened, varying $A_0$ to 300~\GeV{},  $\tan\beta$ to $30$ or $\mu$ to $-\mu$ leads to significant ($\sim 5\%$) changes, among the strongly interacting particles, only in the stop and sbottom masses.
Accordingly, the exclusion limits are not strongly sensitive to these parameters.

\maybesection{8. Summary}

This \note{} reports a search for new physics in final states containing
high-\pT{} jets, missing transverse momentum and no electrons or muons.
Good agreement is seen between the numbers of events observed in the
four signal regions and the numbers of events expected from SM sources. 
Signal regions A, B, C and D exclude non-SM cross sections within acceptance of
1.3,    
0.35,   
1.1     
and
0.11~pb 
respectively at 95\% confidence.  

The results are interpreted in both a simplified model containing
only squarks of the first two generations, a gluino octet and a
massless neutralino, as well as in MSUGRA/CMSSM models with $\tan\beta=3$, $A_0=0$
and $\mu>0$. In the simplified model, gluino masses below $500$~GeV
are excluded at the $95\%$ confidence level with the limit increasing
to $870$ GeV for equal mass squarks and gluinos. In the MSUGRA/CMSSM
models equal mass squarks and gluinos below $775$~GeV are excluded. 
  
%
\maybesection{9. Acknowledgements}

We wish to thank CERN for the efficient commissioning and operation of the LHC during this initial high-energy data-taking period as well as the support staff from our institutions without whom ATLAS could not be operated efficiently.

We acknowledge the support of ANPCyT, Argentina; YerPhI, Armenia; ARC, Australia; BMWF, Austria; ANAS, Azerbaijan; SSTC, Belarus; CNPq and FAPESP, Brazil; NSERC, NRC and CFI, Canada; CERN; CONICYT, Chile; CAS, MOST and NSFC, China; COLCIENCIAS, Colombia; MSMT CR, MPO CR and VSC CR, Czech Republic; DNRF, DNSRC and Lundbeck Foundation, Denmark; ARTEMIS, European Union; IN2P3-CNRS, CEA-DSM/IRFU, France; GNAS, Georgia; BMBF, DFG, HGF, MPG and AvH Foundation, Germany; GSRT, Greece; ISF, MINERVA, GIF, DIP and Benoziyo Center, Israel; INFN, Italy; MEXT and JSPS, Japan; CNRST, Morocco; FOM and NWO, Netherlands; RCN, Norway;  MNiSW, Poland; GRICES and FCT, Portugal;  MERYS (MECTS), Romania;  MES of Russia and ROSATOM, Russian Federation; JINR; MSTD, Serbia; MSSR, Slovakia; ARRS and MVZT, Slovenia; DST/NRF, South Africa; MICINN, Spain; SRC and Wallenberg Foundation, Sweden; SER,  SNSF and Cantons of Bern and Geneva, Switzerland;  NSC, Taiwan; TAEK, Turkey; STFC, the Royal Society and Leverhulme Trust, United Kingdom; DOE and NSF, United States of America.  

The crucial computing support from all WLCG partners is acknowledged gratefully, in particular from CERN and the ATLAS Tier-1 facilities at TRIUMF (Canada), NDGF (Denmark, Norway, Sweden), CC-IN2P3 (France), KIT/GridKA (Germany), INFN-CNAF (Italy), NL-T1 (Netherlands), PIC (Spain), ASGC (Taiwan), RAL (UK) and BNL (USA) and in the Tier-2 facilities worldwide.


\bibliographystyle{atlasnote}
\bibliography{PUB_0lepton2010}

\clearpage
\begin{flushleft}
{\Large The ATLAS Collaboration}

\bigskip

G.~Aad$^{\rm 48}$,
B.~Abbott$^{\rm 111}$,
J.~Abdallah$^{\rm 11}$,
A.A.~Abdelalim$^{\rm 49}$,
A.~Abdesselam$^{\rm 118}$,
O.~Abdinov$^{\rm 10}$,
B.~Abi$^{\rm 112}$,
M.~Abolins$^{\rm 88}$,
H.~Abramowicz$^{\rm 153}$,
H.~Abreu$^{\rm 115}$
% ...
\bigskip

% ...
$^{10}$ Institute of Physics, Azerbaijan Academy of Sciences, Baku, Azerbaijan\\
$^{11}$ Institut de F\'isica d'Altes Energies and Universitat Aut\`onoma  de Barcelona and ICREA, Barcelona, Spain\\
% ...
$^{48}$ Fakult\"{a}t f\"{u}r Mathematik und Physik, Albert-Ludwigs-Universit\"{a}t, Freiburg i.Br., Germany\\
$^{49}$ Section de Physique, Universit\'e de Gen\`eve, Geneva, Switzerland\\
% ...
$^{88}$ Department of Physics and Astronomy, Michigan State University, East Lansing MI, United States of America\\
% ...
$^{111}$ Homer L. Dodge Department of Physics and Astronomy, University of Oklahoma, Norman OK, United States of America\\
$^{112}$ Department of Physics, Oklahoma State University, Stillwater OK, United States of America\\
% ...
$^{115}$ LAL, Univ. Paris-Sud and CNRS/IN2P3, Orsay, France\\
% ...
$^{118}$ Department of Physics, Oxford University, Oxford, United Kingdom\\
% ...
$^{153}$ Raymond and Beverly Sackler School of Physics and Astronomy, Tel Aviv University, Tel Aviv, Israel\\
% ...
\end{flushleft}


\end{document}